\documentclass [%
 reprint,nofootinbib,
superscriptaddress,
 amsmath,amssymb,
 aps,
]{revtex4-1}
\makeatletter
\renewcommand{\fnum@figure}{Fig. \thefigure}

\newcommand{\mcite}[1]{\textcolor{red}{ MISSING CITATION}}
\newcommand{\Msun}{\rm M_{\rm Sun}}
\makeatother

\usepackage{soul}
\usepackage{xcolor}
\usepackage{graphicx}
\usepackage{dcolumn}
\usepackage{bm}
\usepackage[T1]{fontenc}
\usepackage{subcaption}
\usepackage{mwe}
\usepackage{chngcntr}

\begin{document}

\preprint{APS/123-QED}

\title{GW190814 as a massive rapidly-rotating neutron star with exotic degrees of freedom}

\author{V. Dexheimer}
\email{vdexheim@kent.edu}
\affiliation{Department of Physics, Kent State University, Kent, OH 44243 USA}

\author{R.O. Gomes}
\affiliation{Frankfurt Institute for Advanced Studies, 60438 Frankfurt am Main, Germany}

\author{T. Kl\"ahn}
\affiliation{Department of Physics and Astronomy, California State University Long Beach, Long Beach, CA 90840, USA}

\author{S. Han}
\affiliation{Department of Physics, University of California, Berkeley, CA~94720, USA} 
\affiliation{Department of Physics and Astronomy, Ohio University, Athens, OH~45701, USA}

\author{M. Salinas}
\affiliation{Department of Physics and Astronomy, California State University Long Beach, Long Beach, CA 90840, USA}

\date{\today}

\begin{abstract}
In the context of the massive secondary object recently observed in the compact-star merger GW190814, we investigate the possibility of producing massive neutron stars from a few different equation of state models that contain exotic degrees of freedom, such as hyperons and quarks. Our work shows that state-of-the-art relativistic mean field models can generate massive stars reaching $\gtrsim 2.05\,\Msun$, while being in good agreement with gravitational-wave events and x-ray pulsar observations, when quark vector interactions and non-standard self-vector interactions are introduced. In particular, we present a new version of the Chiral Mean Field (CMF) model in which a different quark-deconfinement potential allows for stable stars with a pure quark core. When rapid rotation is considered, our models generate stellar masses that approach, and in some cases surpass $2.5\,\Msun$. We find that in such cases fast rotation does not necessarily suppress exotic degrees of freedom due to changes in stellar central density, but require a larger amount of baryons than what is allowed in the non-rotating stars. This is not the case for pure quark stars, which can easily reach $2.5\,\Msun$ and still possess approximately the same amount of baryons as stable non-rotating stars. We also briefly discuss possible origins for fast rotating stars with a large amount of baryons and their stability, showing how the event GW190814 can be associated with a star containing quarks as one of its progenitors.
\end{abstract}                       
                              
\maketitle


\section{Introduction}

In the last decade, several massive neutron stars have been observed, $1.97 \pm 0.04\,\Msun$ \cite{Demorest:2010bx}, $2.01 \pm 0.04\,\Msun$ \cite{Antoniadis:2013pzd}, $2.27 \pm 0.17\,\Msun$ \cite{Linares:2018ppq}, and $2.14^{+0.20}_{-0.18}\,\Msun$ \cite{Cromartie2020}, all consistent with having a mass of $\sim 2.1\,\Msun$. While some of these observed neutron stars are rapidly spinning, we note that the most rapid known pulsar PSR J1748-2446ad was found to rotate with a frequency of $716$ Hz \cite{Hessels:2006ze}. Recently, however, there has been indication of even faster stars, such as one possibly rotating at $1250$ Hz inferred from observations of narrow pulses in the fast radio burst FRB 181112 \cite{Yamasaki:2020cqo}.

Following the multi-messenger gravitational wave event GW170817 \cite{GW170817}, it was argued that the maximum mass of neutron stars is likely be bounded by $M \lesssim 2.3\, \Msun$ \cite{Rezzolla2018}. It has been shown \cite{Rezzolla2018} that such an upper bound would allow for uniformly rotating stars of up to $M\lesssim 2.8\,\Msun$. Recently, the LIGO/VIRGO collaboration has announced the gravitational wave event GW190814 \cite{Abbott:2020khf}, which was reported to be the merger of a $23.2 ^{+1.1}_{-1.0}\,\Msun$ black hole and a $2.59^{+0.08}_{-0.09}\,\Msun$ object. The secondary object's mass falls into the so called ``mass-gap'' category, in which stars are considered too light to be a black hole but too heavy to be a neutron star, the latter being due to lack of electromagnetic observations, but also due to conflicts with our current knowledge of supernova explosion mechanisms \citep{Fryer2012}. See  Ref.~\cite{Zevin:2020gma} for a recent discussion on the mass-gap in the context of GW190814, where the authors state that such objects can be formed, but at very low rates. See also Ref.~\citep{Safarzadeh:2020ntc} for an alternative explanation in the context of accretion from the circumbinary disk, which is only possible in the case of large mass companions. In this case, the fast rotation from the neutron star could be supplied by the circumbinary accretion disk. 

In the case of dynamical exchanges in dense stellar environments, the merger rate of neutron stars with black holes may be significant in young star clusters \cite{10.1093/mnras/stu824}, with the possibility of the lower-mass object being itself a merger remnant that acquires a black hole companion via dynamical interactions \cite{Gupta:2019nwj}. In the latter case, the high spin of the lower-mass object would be the distinguishing feature. Unfortunately,  the uninformative spin posterior
for the secondary object in GW190814 provides no evidence for or against this hypothesis \cite{GW170817}. Hopefully, in the near future, this matter could be settled through the indication or not of a second burst
of mass ejection in similar (mass-wise) mergers \cite{Most:2020exl}.

Recently, it was found that non-rotating nucleonic models can generate massive stars with $M\sim2.5\,\Msun$
\cite{Tan:2020ics,Tsokaros:2020hli,Lim:2020zvx,Huang:2020cab,Rather:2020lsg}, even when allowing for kaon condensation \cite{Thapa:2020usm}, although it was shown that some of these models are not necessarily compatible with constraints obtained from energetic heavy-ion collisions \cite{Fattoyev:2020cws}. Including fast rotation, Ref.~\cite{Zhang:2020zsc} demonstrated that using a parametrized nucleonic equation of state (EoS), $M>2.5\,\Msun$ stars can be stable even though, again, this can create tension with other astrophysical observables \cite{Tews:2020ylw}. And, although it was shown that rotation can increase the mass of hybrid stars, in Refs.~\cite{Shahrbaf:2019vtf,Dhiman:2007ck,2020arXiv201203232R} it was found that only nucleonic rotating stars can reach $M>2.5\,\Msun$. In addition, more recently, Ref.~\cite{Sedrakian:2020kbi,Li:2020ias} concluded that, even when including rotation, the presence  of hyperons or heavier  resonances in dense matter acts strongly against the interpretation of GW190814 involving a neutron star.

Following recent advances in the literature showing that massive neutron stars should contain a quark core \cite{Annala:2019puf}, in this work, we investigate the possibility of having exotic degrees of freedom in rapidly rotating neutron stars. We apply realistic relativistic models to describe the interior of massive neutron stars as containing hyperons \cite{Glendenning:1982nc} and/or quarks \cite{Itoh:1970uw,Glendenning:1992vb}, which is possible due to the introduction of different vector (repulsive) interactions. Although some of these interactions have already individually been introduced, in this work we present a more in-depth discussion of their effects. In particular, we discuss the introduction of quark-vector interactions and free non-standard self-vector interactions to the Chiral Mean Field (CMF) model. These, combined with a different quark-deconfinement potential, allows for stable stars with a pure quark core to be described for the first time in this formalism, which is also able to describe at finite temperature lattice QCD and heavy-ion collisions data.

Higher-order vector interactions such as $\omega^4$ have been suggested long ago and first used in Walecka-type models \cite{Bodmer:1991hz}. For quark matter, this kind of interaction was recently used in Ref.~\cite{Lopes:2020iib} for the bag model. Quark vector-isoscalar and vector-isovector couplings have been used in the NJL model \cite{Baym:2017whm,Pereira:2016dfg,Wu:2018kww,Malfatti:2019tpg,Shahrbaf:2019vtf,Lopes:2020rqn,Ferreira:2020evu,Alaverdyan:2020xnv}, 2-flavor confining QM (CQM) model \cite{Cao:2020zxi}, FRG within a quark-meson model \cite{Otto:2020hoz}, quark-meson-nucleon (QMN) model \cite{Marczenko:2020jma}, and bag model \cite{Lopes:2020iib}.  Finally, a 8-quark interaction has been used in the NJL model \cite{Benic:2014jia}. All of these have been shown to generate more massive neutron-stars than their respective zero-vector coupling counterparts

In a second step, we introduce fast rotation effects using a full general relativity numerical code RNS \cite{Stergioulas95}. In this way we are able to verify how rotation changes the stars we generate by allowing them to hold more mass, but at the same time present a different internal structure. For a detailed discussion on how uniform and differential rotation changes the masses and central densities of hybrid stars, see Ref.~\cite{Bozzola:2019tit}, where the authors find massive fast rotating stars with M$\sim2.5$ $\Msun$ using a constant speed of sound parametrization with $v_s=0.8$ c.

According to Ref.~\cite{Biswas:2020xna}, fast rotation larger than $1.1$ kHz is possible, but in this case  instabilities related to f-modes and r-modes have to be somehow suppressed. This can be possible due to damping caused by mutual friction of superfluid vortices \cite{Lindblom:1999wi,Gaertig:2011bm,10.1111/j.1365-2966.2009.14963.x}, specially in the case of matter with hyperons and quarks \cite{PhysRevD.73.084001}. Alternatively, instabilities could still be damped if the temperature stays below a certain rotation dependent threshold. According to Ref.~\cite{Zhou:2020xan}, this mechanism alone could explain a $2.50-2.67$ $\Msun$ neutron star.

Our goal is to verify to which extent (if any) fast rotation prevents exotic degrees of freedom from being present in the massive hybrid stars we obtain. We finalize by discussing how our results compare to results obtained assuming that fast rotating massive stars are completely made out of quarks.

\section{Results}

The intermediate and high density regimes ($\gtrsim2$ $n_0$, the nuclear saturation density) that occupy a significant portion of the volume inside neutron stars are regions where neither of the reliable theories of Effective Field Theory for nucleons at low density or Perturbative Quantum Chromo Dynamics (PQCD) at extremely high density can be directly applied. As a result, there are very few options left. One of them is to resort to some sort of interpolation \cite{Kurkela2014, Most2018} and another is to rely on relativistic mean-field effective models. In this work, we make use of the latter, which unlike the former, can provide a particle population and, therefore, can be tested in dynamical stellar simulations of, for example,
stellar cooling. 

\subsection{The CMF Model}

In this subsection, we focus on the Chiral Mean Field (CMF) model. We briefly describe its formalism and present three modifications that allow it to describe for the first time stable hybrid stars with a pure quark core.  The CMF model is based on a nonlinear realization of the SU(3) sigma model and is constructed in such a way that chiral invariance is restored at large temperatures and/or densities. In its present version, it contains hadronic, as well as quark degrees of freedom \footnote{Note that an alternative version of the CMF model includes in addition the chiral partners of the baryons and gives the baryons a finite size \cite{Steinheimer:2011ea, Motornenko:2019arp}}. The masses of the baryons and quarks are generated by the medium and are also coupled to a field $\Phi$, which acts as an order parameter for deconfinement: 
\begin{eqnarray}
M_{B}^* &= g_{B \sigma} \sigma + g_{B \delta} \tau_3 \delta + g_{B \zeta} \zeta + M_{0_B} + g_{B \Phi} \Phi^2,  \nonumber \\
M_{q}^* &= g_{q \sigma} \sigma + g_{q \delta} \tau_3 \delta + g_{q \zeta} \zeta + M_{0_q} + g_{q \Phi}(1 - \Phi).
\end{eqnarray}
In this way, at low densities (and or temperatures) the quarks are too massive to appear, while at large densities (and or temperatures) the baryons become too massive and disappear. At zero temperature, $\Phi$ jumps from $0\to\sim1$ across the phase transition. The scalar mesons $\sigma$, $\delta$ (isovector), and $\zeta$ (with closed strangeness) and the vector mesons $\omega$, $\rho$ (isovector), and $\phi$ (with closed strangeness) mediate the interactions between the baryons and quarks. The mesons are taken as classical fields within the mean field approximation \cite{Chin:1974sa}. See Ref.~\cite{Roark:2018uls} for a complete list of couplings $g$ and bare masses $M_0$.

\begin{figure}[t!]
\includegraphics[{trim=0.5cm 0 0 2.4cm},clip,width=9.4 cm]{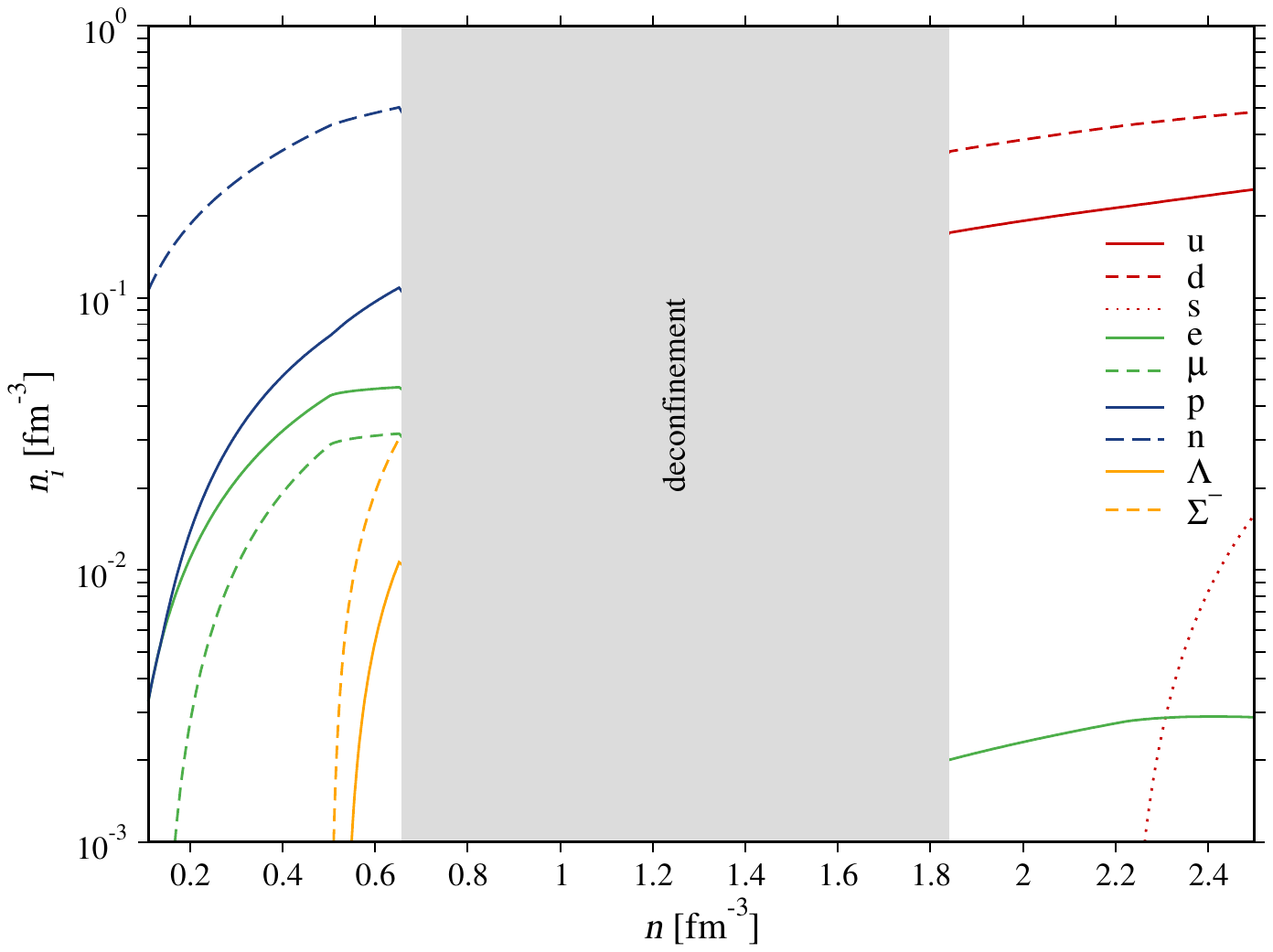}
\includegraphics[{trim=0.5cm 0 0 2.4cm},clip,width=9.4 cm]{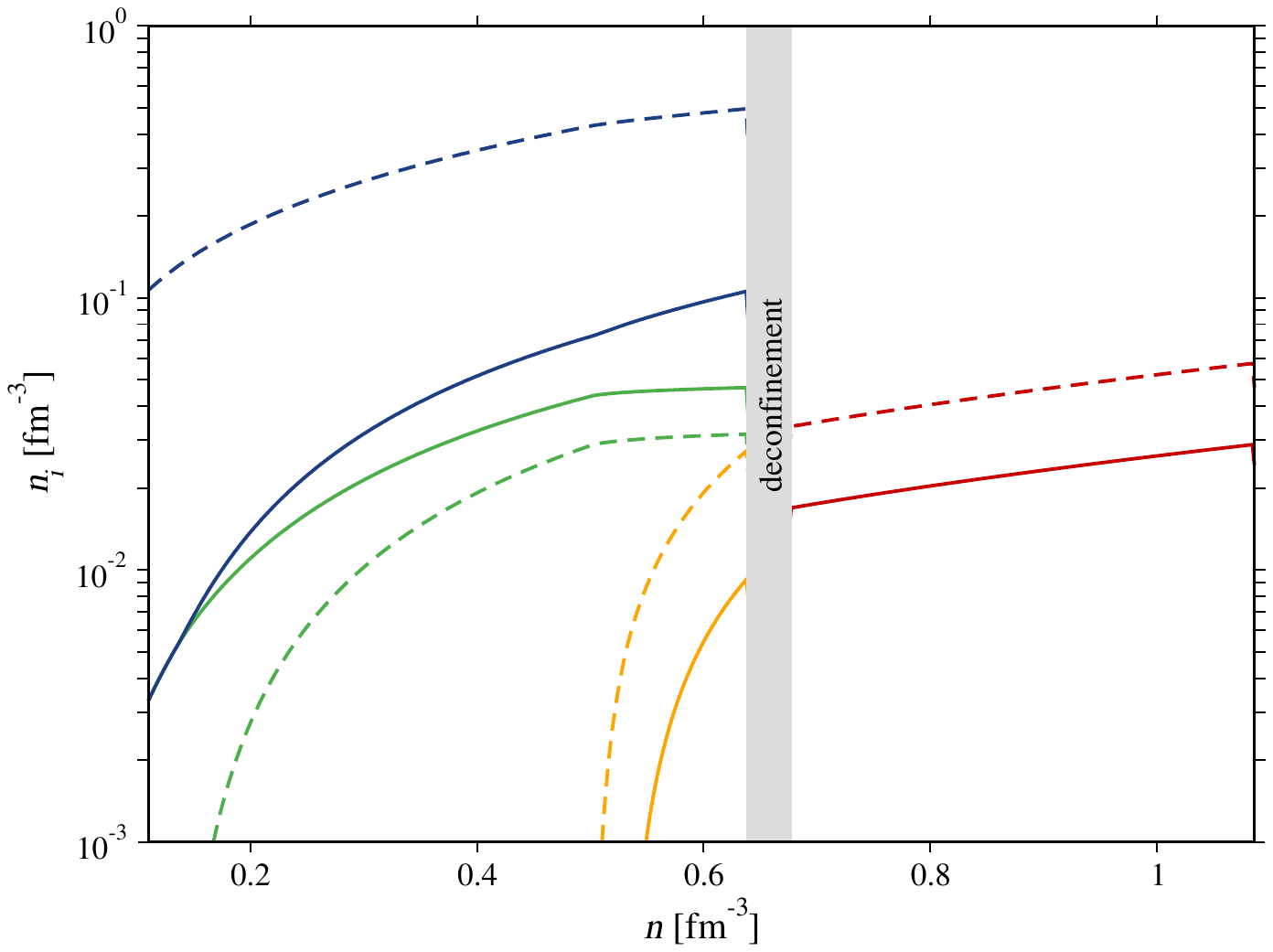}
\caption {Particle population within the CMF model with quarks as a function of baryon number density. The shaded region covers the jump in density generated by the first-order phase transition to deconfined quark matter. Quark number densities were divided by $3$. The top panel shows the original parametrization with an additional $\omega\rho$ interaction. The bottom panel shows the effects of a modified $U$ potential and a finite quark vector-meson coupling. In both cases we keep $\omega^4=\omega^6=0$.}
\label{fig1}
\end{figure}

The Lagrangian density of the model:
\begin{eqnarray}
\mathcal{L} = \mathcal{L}_{\rm{Kin}} + \mathcal{L}_{\rm{Int}} + \mathcal{L}_{\rm{Self}} + \mathcal{L}_{\rm{SB}} - U ,
\end{eqnarray}
contains a kinetic term for the baryons, quark, leptons, and mesons, an interaction term between baryons and or quarks mediated by the mesons, a self-interaction term for the mesons, a symmetry breaking term responsible for producing the masses of the pseudo-scalar mesons, and a potential for the deconfinement order parameter $\Phi$:
\begin{eqnarray}
U &=& \big(a_o T^4 + a_1 \mu_B^4 + a_2 T^2 \mu_B^2 \big) \Phi^2 \nonumber \\
&+& a_3 T_o^4 \ \ln{\big(1 - 6 \Phi^2 + 8 \Phi^3 -3 \Phi^4 \big)} ,
\label{upol}
\end{eqnarray}
which depends on temperature $T$ and baryon chemical potential $\mu_B$. For details on the other $U$ terms and values for the couplings $a$, please see Ref.~\cite{Roark:2018uls}. The chemical potential dependency in $U$ (still present even at zero temperature) was introduced motivated by the discussions in Ref.~\cite{Schaefer:2007pw}.

The CMF model was fitted to reproduce low and high-energy nuclear and lattice QCD constraints \cite{Dexheimer:2009hi}, such as the deconfinement crossover transition expected to take place at very large temperatures, as well as tested in stellar merger \cite{Most:2018eaw} and cooling simulations \cite{Negreiros:2010hk}. For a detailed description of how this formalism can be applied to describe neutron and proto-neutron stars, while being in agreement with perturbative QCD results for the relevant regime, see Ref.~\cite{Roark:2018uls,Roark:2018boj}.

More recently, we have introduced a new free parameter in the meson self-interaction term of the pure hadronic CMF model \cite{Dexheimer:2018dhb}, a vector-isovector coupling proportional to $\omega^2\rho^2$. It was shown that this kind of term, first introduced in Ref.~\cite{Horowitz:2002mb} to improve agreement with neutron skin data, allows us to describe a more soft EoS at low/intermediate densities and, as a consequence, reproduce stars with smaller radius and lower tidal deformability (with practically unchanged mass), in addition to being in better agreement with Effective Field Theory results for low densities \cite{Hebeler:2013nza}. We set our normalized $\omega\rho$ coupling strength here to $11.34$ in order to generate tidal deformabilities $\tilde{\Lambda}<730$, in agreement with results obtained from the binary neutron-star merger GW170817 \citep{LIGOGW170817}. The $\rho$ coupling constant is then refit to reproduce the same symmetry energy at saturation (as the original CMF parametrization), $g_{N\rho}=4.41$, although it generates now a lower symmetry energy slope $L=75$ MeV in better agreement with data (for a discussion on the symmetry energy see Ref.~\cite{Kolomeitsev:2016sjl} and references therein).

Our results for the CMF model at zero temperature under charge neutrality and chemical equilibrium with the $\omega\rho$ interaction are shown (for the first time with deconfinement to quark matter) in the top panel of Fig.~1 as a function of baryon number density. When compared to the original CMF parametrization (not shown here), this one generates matter which is more soft symmetry energy, meaning a lower energy cost to produce neutrons and, therefore, a larger neutron-to-proton ratio, less leptons and hyperons. Note that, alternatively, varying the scalar-isoscalar coupling strength would produce a similar outcome \cite{Kubis:2020ysv}, although this change is not allowed in chiral models (such as ours), in which the scalar sector is fixed in order to generate the vacuum masses of hadrons.

The $\omega\rho$ interaction also reproduces a slightly later (with respect to baryon chemical potential or density) deconfinement phase transition. The quark phase stays unchanged, as the quarks do not couple in the isovector channel. The delay of the phase transition has to do with a lower pressure vs. energy density (softer EoS) at low/intermediate densities but that corresponds to a stiffer pressure vs. baryon chemical potential in the hadronic phase when the $\omega\rho$ interaction is included. The gray lines in Fig.~2 illustrate how the stellar radius is modified by varying the $\omega\rho$ interaction (shown for hadronic matter only).

\begin{figure}[t!]
\includegraphics[{trim=0.5cm 0 0 2.4cm},clip,width=9.4 cm]{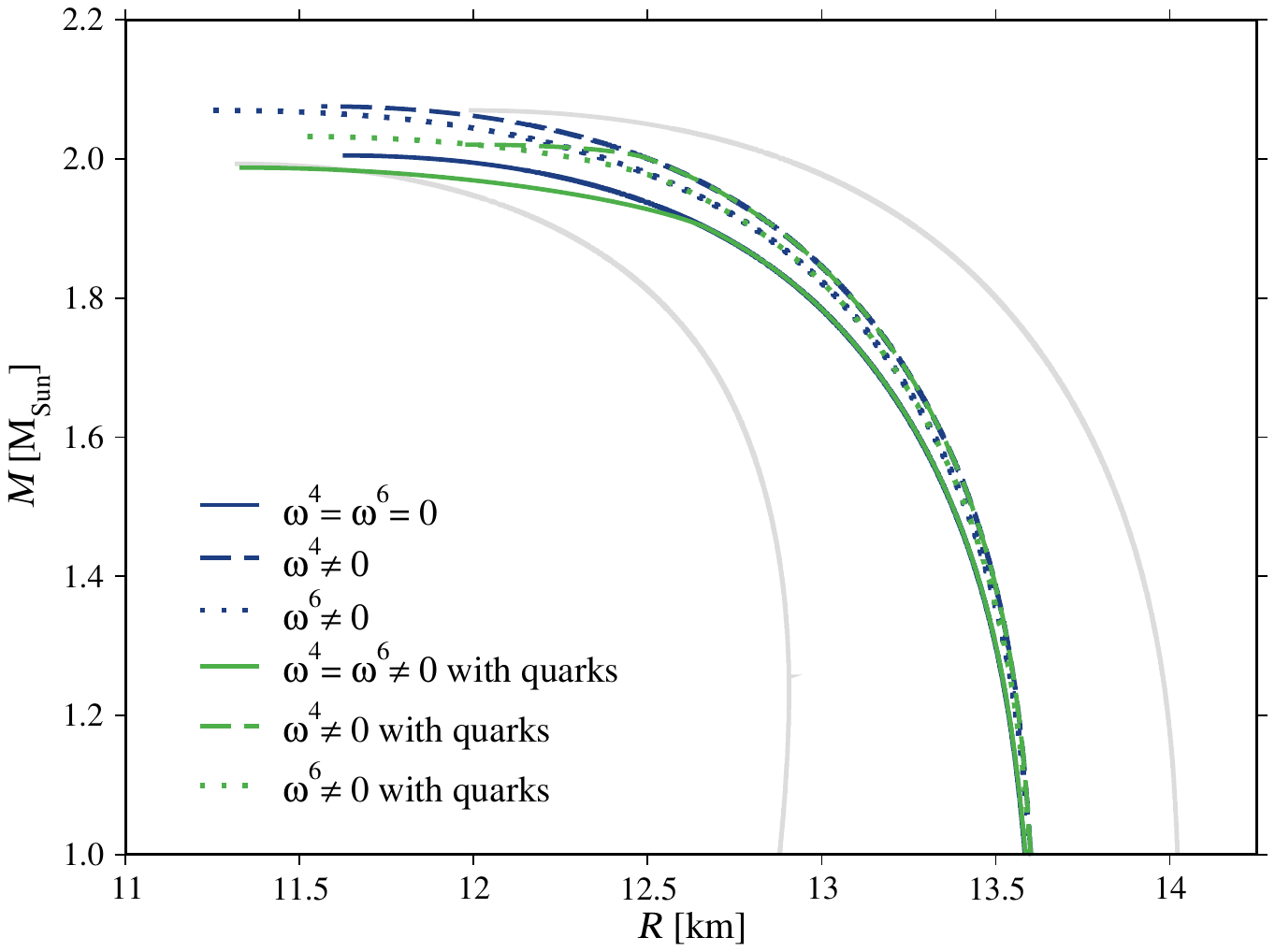}
\caption {Static mass-radius diagram for several CMF model equations of state (shown until the maximum mass only). Green curves present a first-order phase transition to deconfined quark matter. The grey lines show the results of varying the $\omega\rho$ vector-isovector self interaction coupling strength from $0\to62$ for hadronic matter.}  
\label{fig1}
\end{figure}

\begin{table}[]
\begin{tabular}{c|ccccc}
~Interactions~    &~M($\Msun$)~&~$n_c(n_0)$~&~$v_s^2$~&~$R_{1.4}$~(km)~&~$\Lambda_{1.4}$~\\ 
\hline
hadrons only\\ 
    $\omega^4=\omega^6=0$   &   2.00   &  6.5   &   0.47   &   13.4   &  711  \\
    $\omega^4\ne0$          &   2.08   &  6.5   &   0.57   &   13.5   &  730  \\
    $\omega^6\ne0$          &   2.07   &  6.8   &   0.96   &   13.5   &  722  \\
\hline 
with quarks\\ 
    $\omega^4=\omega^6=0$   &   1.99   &  6.9   &   0.60   &   13.4   &  711  \\ 
    $\omega^4\ne0$          &   2.02   &  6.0   &   0.53   &   13.5   &  730  \\ 
    $\omega^4\ne0$ $*$      &   2.07   &  6.3   &   0.56   &   13.5   &  730  \\
    $\omega^6\ne0$          &   2.03   &  6.6   &   0.61   &   13.5   &  722  \\
    $\omega^6\ne0$ $*$      &   2.07   &  6.7   &   0.61   &   13.5   &  722  \\
\end{tabular}
\begin{center}
\caption{For each different set of high-order vector self interactions, we show the maximum allowed stellar mass $M$ of the static stellar sequence, its central density $n_c$ (in units of saturation number density $n_0=0.15$ fm$^{-3}$), its central speed of sound squared $v_s^2$, and the radius $R$ and tidal deformability $\Lambda$ for a $M=1.4\,\Msun$ star. The top rows show results for hadronic (only) matter and the bottom when allowing for a phase transition to deconfined quark matter. The results marked by an asterisk reproduce a late (very high density) phase transition to quark matter.
} 
\end{center}
\label{tab1}
\end{table}

But, as a consequence of the large jump in baryon density across the deconfinement phase transition shown in the top panel of Fig.~1, stars with pure quark matter inside (that have reached the threshold for deconfinement to quark matter to take place) are not stable. Of course, this is not the case if we allow for a mixture of phases to exist, which can turn hybrid stars stable \cite{Roark:2018boj}, but this is not the topic of the present work.

In this work, we present three additional modifications to the CMF model, all of which will allow to describe massive hybrid stars with a pure quark core. They are:
\begin{itemize}
\item a change in the deconfinement order parameter potential U, which now has a weaker quadratic dependence on the pure chemical potential part $a_1\textquoteright \mu_B^2 \Phi^2$, instead of the quartic one in Eq.~3, with $a_1\textquoteright=-2867.5$. This change requires an adjustment of the $\Phi$ coupling in the effective mass of the particles in Eq.~(1) to $g_{q\Phi}=450$ and $g_{B\Phi}=1350$ MeV. As a consequence, the deconfinement first-order phase transition becomes much weaker, generating a much smaller jump in density. This can be verified when comparing both panels in Fig.~1. Note that the bottom panel of Fig.~1 and all following particle population figures shown in this work will extend only to density ranges that go just a little bit beyond the maximum density reached in maximum mass static stars. As $\Phi$ is zero in the hadronic phase at zero temperature, the changes we discussed in $U$ and $\Phi$ couplings affect only the quark phase.
\item a change in the quark couplings in the interaction term of the Lagrangian density (which contains the effective masses $M^*$). We have decreased $g_{q \sigma}$ and $g_{q \zeta}$ to $-0.5$ and, for the first time within the CMF model, added a non-zero vector-quark interaction with coupling $g_{q \omega}=11$. Together, these changes allow for massive stable stars with a pure quark core. We show in the full lines of Fig.~2 the mass-radius diagram for the population showed in the bottom panel of Fig.~1, also including a Baym, Pethick, and Sutherland (BPS) prescription for the crust \cite{bps}. In Fig.~2 we also show using a different color (navy blue) equations of state in which the quarks were artificially suppressed. In this case, as expected, larger stellar masses are achieved. In any case, this parametrization generates stars with masses up to ~2 $\Msun$. Table I illustrates some properties for the maximum-mass star and $1.4\,\Msun$ star of each sequence.
%
\begin{figure}[t!]
\includegraphics[{trim=0.5cm 0 0 2.4cm},clip,width=9.4 cm]{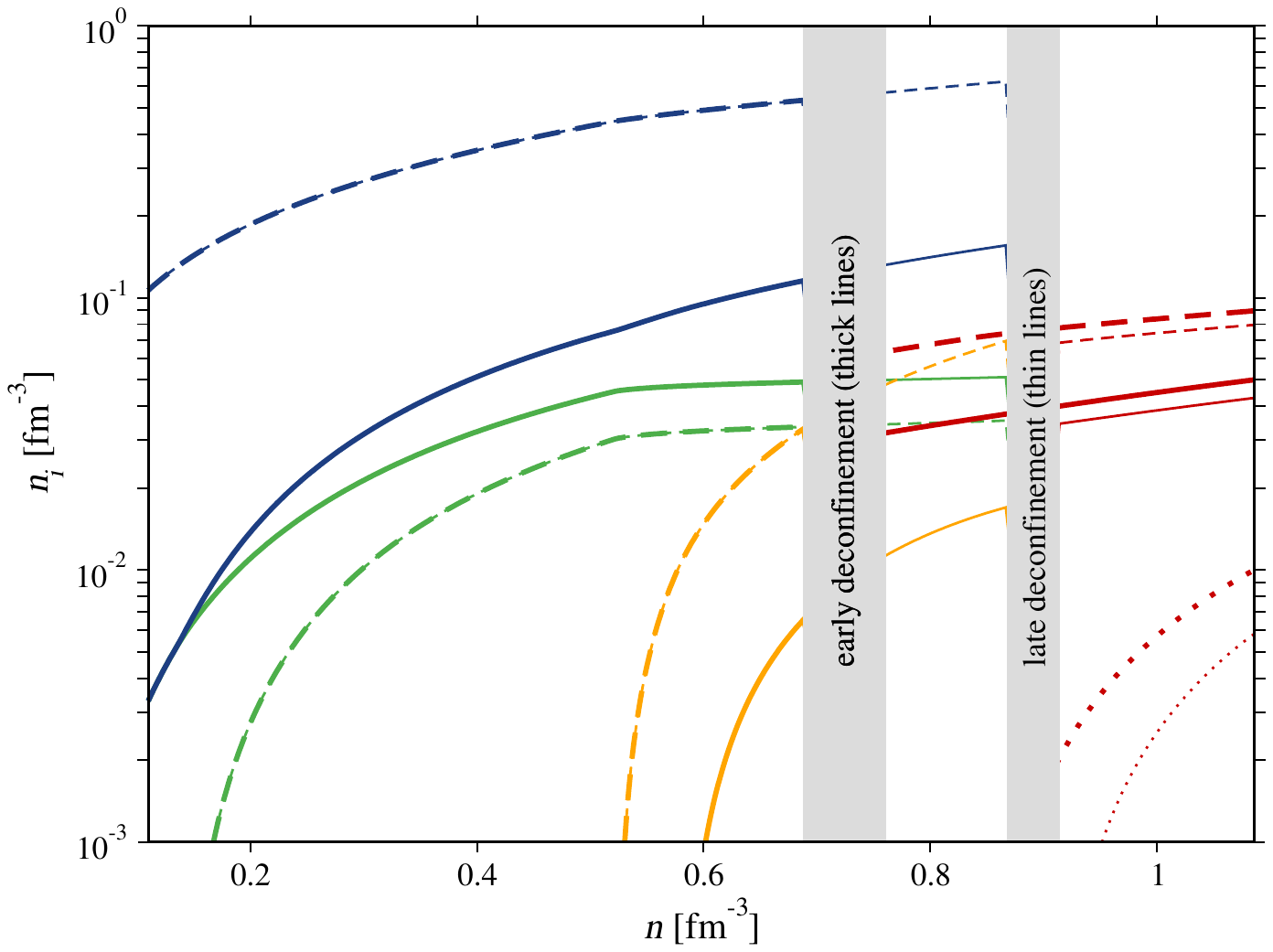}
\includegraphics[{trim=0.5cm 0 0 2.4cm},clip,width=9.4 cm]{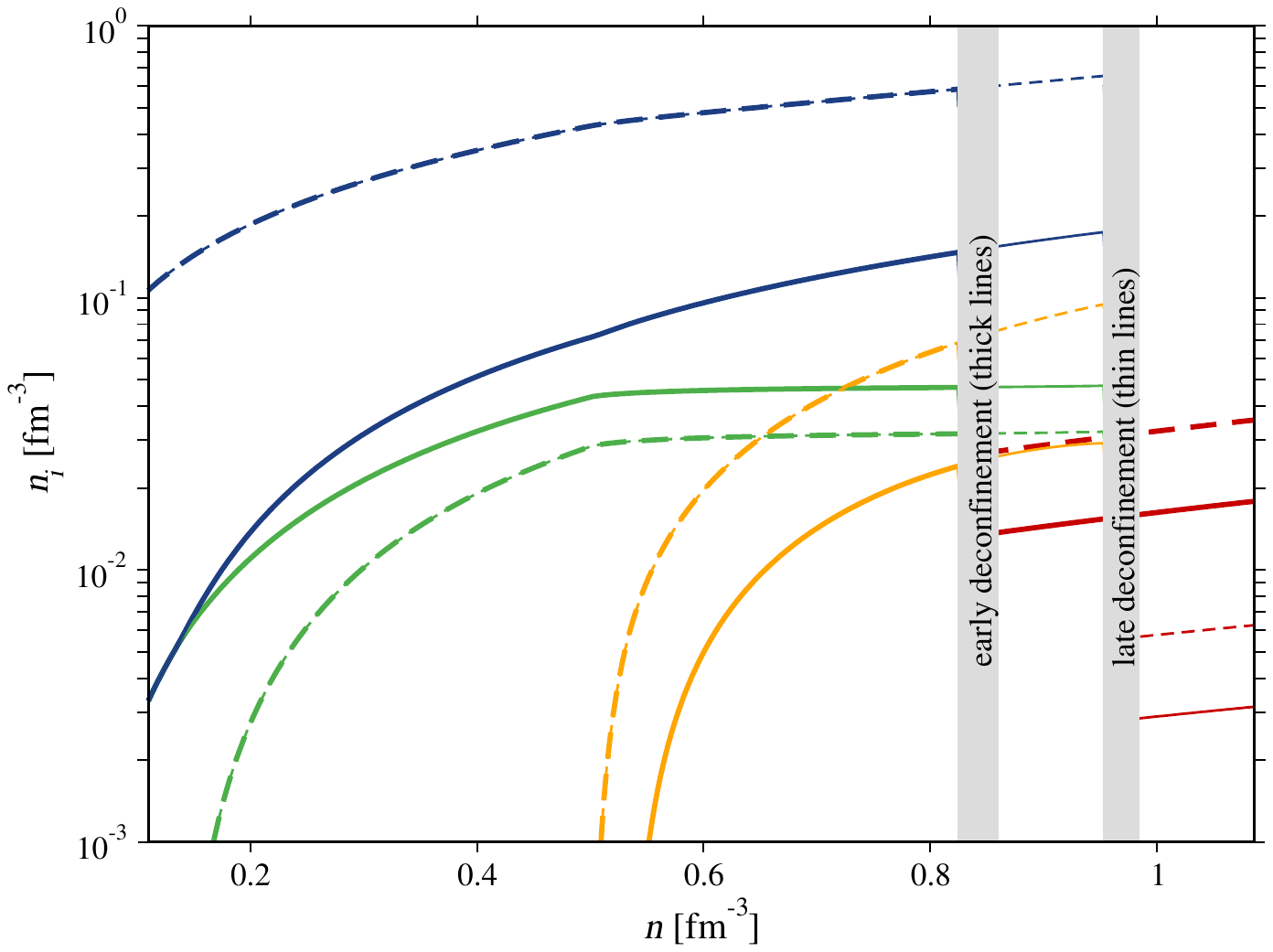}
\caption {Same as the bottom panel of Fig.~1 but with different high order vector self interaction $\omega^4\ne0$ (top panel) and $\omega^6\ne0$ (bottom panel). The thick lines show a parametrization for quark matter that gives rise to an early deconfinement, while the thin lines show a parametrization for quark matter that gives rise to a late deconfinement to quark matter.}
\label{fig1}
\end{figure}

\item a new free parameter in the meson self-interaction term in the Lagrangian. We add either a $\omega^4$ or a $\omega^6$ higher-order vector isovector interaction as a way to take the
Dirac sea into account without resorting to a more complex relativistic Hartree or Hartree-Fock approximation \cite{Furnstahl:1992nu,Furnstahl:1996zm}. These terms allow us to reproduce a stiffer EoS and generate more massive neutron stars (see Fig.~2). In particular, the $\omega^4$ interaction modifies the equation of state at all densities. As a consequence, we modified our original couplings $g_{N\omega}=11.91$ and $g_4=45.20$ to reproduce the original saturation properties, especially being careful not to increase the nuclear compressibility, which in turn constrained it to be larger than our chosen value of $g_{\omega^4}=-4.7$. We changed in this case some quark-related couplings to be $g_{q\omega}=8$, $a_1\textquoteright=-2738$, $g_{q\Phi}=500$, and $g_{B\Phi}=1500$ MeV, in order to warrant stellar stability. We also consider in this work a later (in baryon density or chemical potential) phase transition to quark matter by changing further $g_{q\omega}=8.9$ and $a_1\textquoteright=-2730.6$. This allows us to reproduce even more massive stars, not shown in Fig.~2 because they mimic pure hadronic stars. These late-transition configurations are marked by an asterisk in Table I and also shown in the thin lines in top panel of Fig.~3. The thick lines in this panel show a case that presents an earlier phase transition. Note that the $\omega^4$ interaction allows for the first time for strange quarks to exist inside massive stars in this formalism, appearing in a larger amount in the earlier transition case. It is interesting to note that the $\omega^4$ coupling increases the tidal deformability of $M=1.4\,\Msun$ stars, as well as the central speed of sound of pure hadronic stars, but actually decreases the speed of sound of hybrid stars (see Table~1).

When we add a $\omega^6$ higher-order vector isovector interaction (instead of the $\omega^4$ one), the results differ. This term reproduces a stiffer EoS mainly at large densities. As a consequence, once more we modified our original coupling to $g_4=39.10$ to reproduce the original saturation properties. But, now, our value of $g_{\omega^6}=-0.00038$ MeV$^{-2}$ is chosen to control the speed of sound not to approach the speed of light value too fast with increasing density \footnote{Note that the speed of light as a boundary for the speed of sound is not guaranteed in relativistic models when vector interactions are added \cite{Zeldovich:1962emp}. Usually, this is not a problem, due to the presence of scalar interactions with opposite behavior. This is no longer the case for large densities when the $\omega^6$ self-vector interaction is introduced.}. We changed in this case some couplings to $g_{q\omega}=14.5$, $a_1\textquoteright=-2867.5$, $g_{q\Phi}=500$, and $g_{B\Phi}=1500$ MeV, in order to warrant stellar stability. We again consider a later (in density or chemical potential) phase transition to quark matter by changing further $g_{q\omega}=36$ and $a_1\textquoteright=-2932.25$. This allows us again to reproduce even more massive stars, with configurations marked by an asterisk in Table I and also shown in the thin lines in the bottom panel of Fig.~3. The thick lines in this panel show the earlier phase transition. In either case, with the $\omega^6$ higher-order vector interaction we reproduce stars with more hyperons but no strange quarks.
\end{itemize}

\begin{figure}[t!]
\includegraphics[width=0.48\textwidth]{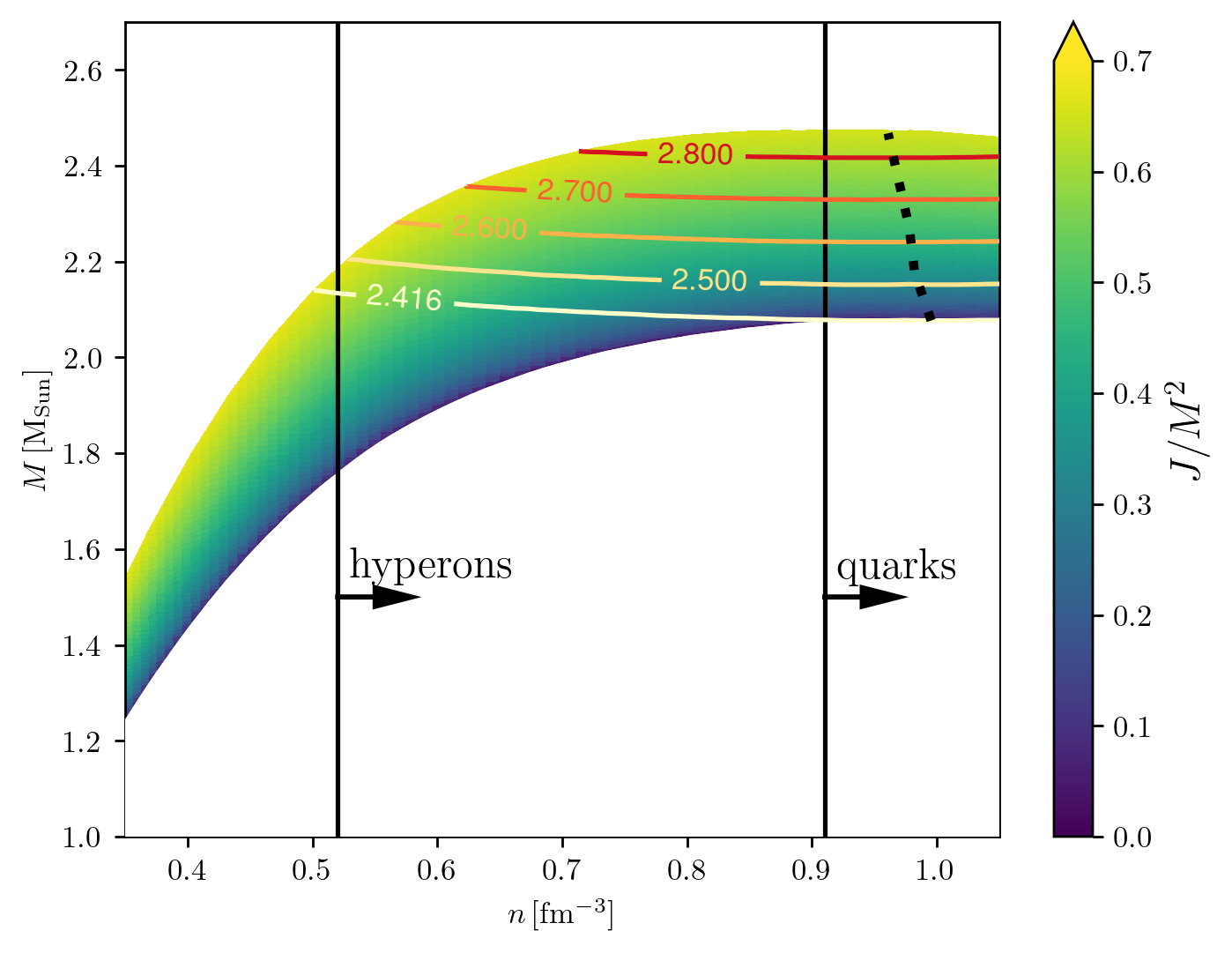}
\includegraphics[width=0.48\textwidth]{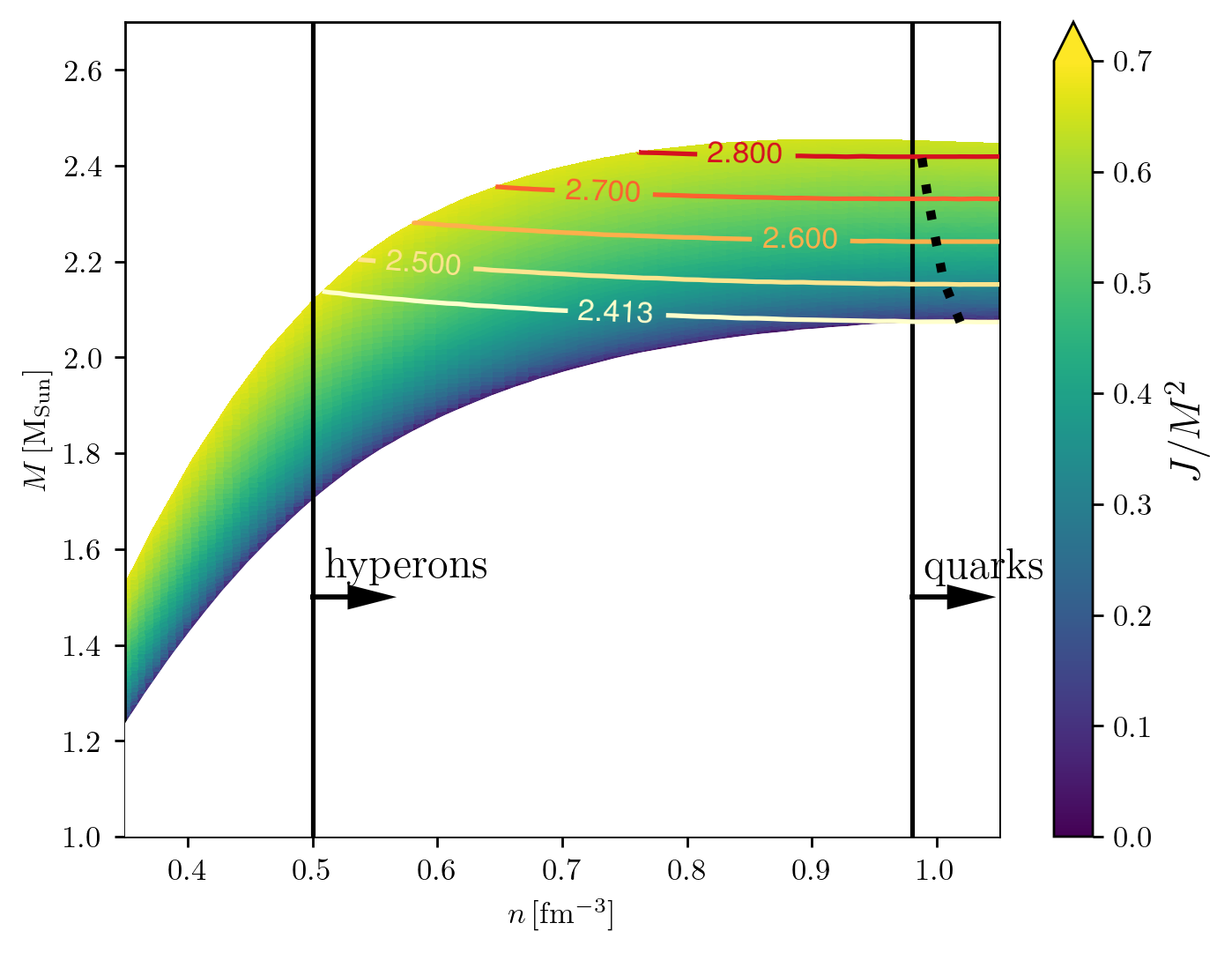}
\caption {Mass-central density diagram for rotating stars reproduced by the CMF model allowing for a late deconfinement to quark matter. The rotational frequency increases vertically and the (almost) horizontal lines denote constant baryon number and are spaced $0.1\,\Msun$ apart. The top panel shows results for the $\omega^4\ne0$ and the bottom panel for the $\omega^6\ne0$ case. The onset of hyperons and quarks is denoted by labelled vertical black lines. The color code shows angular momentum in units of $M^2$. The black dotted lines show the turning-point criterion for stability.}
\label{fig2}
\end{figure}

We generate sequences of rapidly rotating neutron stars using the \texttt{RNS} code \cite{Stergioulas95}, which computes rotating perfect fluid equilibria by numerically fully solving the Einstein equations. Fig.~4 shows how neutron-star gravitational masses increase with increasing rotational frequency within the CMF model for the two most massive cases shown in Table I, including a late phase transition to quark matter and either a $\omega^4$ (top panel) or $\omega^6$ (bottom panel) higher-order self-vector coupling. More specifically, the color code shows angular momentum in units of square mass. Considering for example the most massive static star in either panel of Fig.~4, an increase in angular momentum at fixed baryon number follows the diagonal almost horizontal line, whose effect is to decrease considerably stellar central density, to the point of excluding quarks (both panels) and even hyperons (in the top panel only), when crossing the black threshold vertical lines.

In reality, a fast rotating star does not have to have a static counterpart with the same amount of baryons. In this case however, the star will necessarily collapse to a black hole as it spins down past a given threshold, unless something changes its fate, as for example the merger with another star. Still in this case, the merger of a heavy neutron star with another compact star will result in a black hole, but not before emitting gravitational waves that could be detected here on Earth. This might have been the case of GW190814. As seen in the upper-right corner of both panels in Fig.~4, stars with about $0.4\,\Msun$ extra baryon mass (than the static maximum mass) can have masses $\sim 2.5\,\Msun$ and still contain exotic degrees of freedom, namely hyperons and some quark matter.

\begin{table}[t!]
\begin{tabular}{c|ccc}
~EoS Model~    &~$M$($\Msun$)~&~$f\,{ \rm (kHz)}$~&~ $n_c$ $(n_0)$\\ 
\hline
    CMF $\omega^4\ne0$  H+Q $*$ &  2.474     &  1.38  & 0.88 \\
    CMF $\omega^6\ne0$  H+Q $*$ &  2.455     &  1.39  & 0.89 \\
     MBF + vBag                 &  2.587     &  1.34  & 0.81 \\
     vBag                       &  3.189  & 1.58 & 
     0.71\\    
\end{tabular}
\begin{center}
\caption{Gravitational mass $M$, rotational frequency $f$, and central stellar density $n_c$ for the most massive maximally spinning configuration involved.} 
\end{center}
\label{tab2}
\end{table}

But not all the configurations shown in Fig.~4 are stable. According to the turning-point criterion for secular stability \cite{Friedman:1988er}, moving along a sequence of constant angular momentum, we must find a maximum given by $\partial M/\partial n$. These points are marked by dots, forming the black dotted line in Fig.~4. To the right of this line, uniformly rotating configurations are unstable with respect to axisymmetric perturbations. In the case of the $\omega^4$ self interaction shown in the top panel, there are stable configurations with a considerable amount of quark matter rotating at the maximum allowed frequency and reproducing $\sim 2.5\,\Msun$. On the other hand, in the case of the $\omega^6$ self interaction shown in the bottom panel of Fig.~4, stable configurations have an insignificant amount of quark matter when rotating at the maximum allowed frequency and, therefore, in this case $\sim 2.5\,\Msun$ stars would only contain exotic matter made out of hyperons.
 
For completeness, we report the characteristics of the most massive maximally spinning configuration in the first two lines of Tab.~II.  We note that the maximum mass of the most rapidly spinning hybrid star of the CMF models is about $1.18$ times larger than the mass of the most massive non-spinning star. This is fully in line with universal relations for hadronic stars \cite{Breu2016} but less than what has previously been reported for certain first-order phase transition models \cite{Bozzola:2019tit}.

\subsection{Comparison with other Models}

\begin{figure}[t!]
\includegraphics[{trim=0.5cm 0 0 2.4cm},clip,width=9.4 cm]{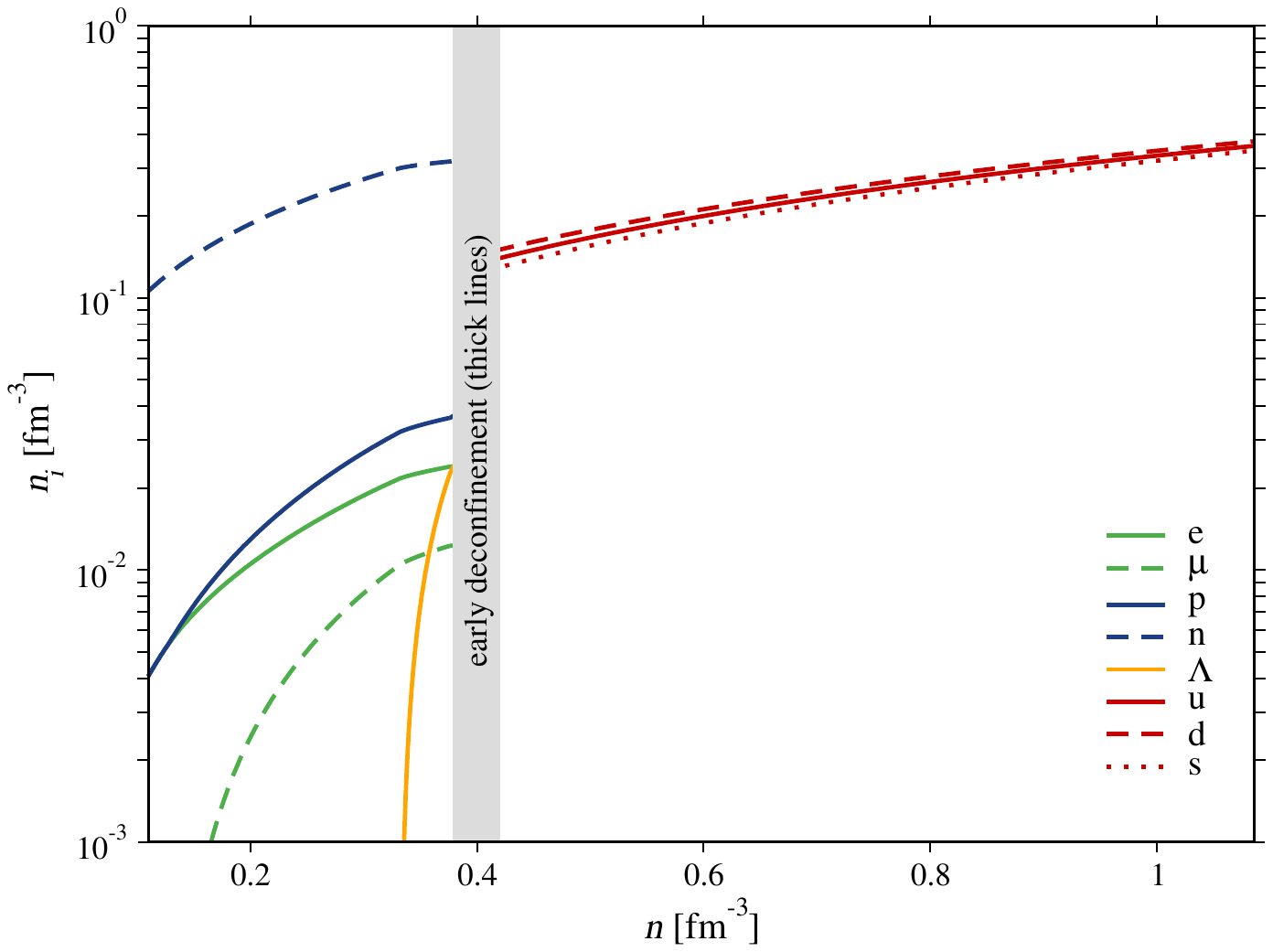}
\includegraphics[{trim=0.5cm 0 0 2.4cm},clip,width=9.4 cm]{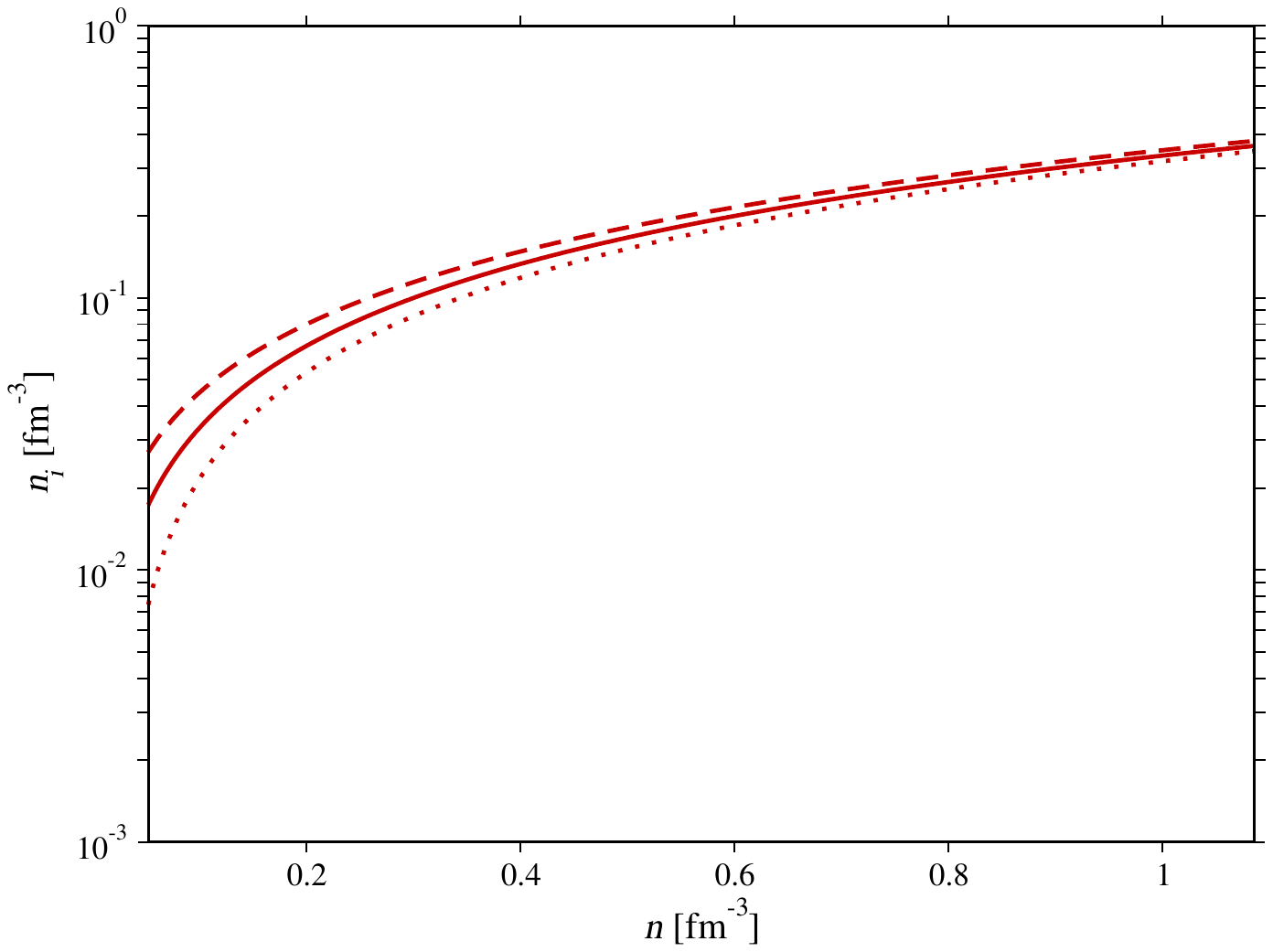}
\caption {Particle population within the MBF-vBag model combination (top panel) and particle population within the vBag model, both as a function of baryon number density. Quark number densities were divided by $3$.} 
\label{fig1}
\end{figure}

For comparison, we also show how vector interactions in the quark EoS can generate massive stars within different models. Note that all of these, including the CMF model, fulfill tidal deformability constraints from Ref.~\citep{LIGOGW170817} and recent NICER x-ray pulsar radius constraints from Refs.~\cite{Miller:2019cac,Riley:2019yda}. We start with a model combination, matching the relativistic Many-body Forces (MBF) model \cite{Gomes:2014aka} which describes the interaction of baryons considering higher-order scalar self-couplings interpreted as many-body forces (parameter $\zeta=0.04$), with a vector-enhanced Bag (vBag) model. We once more add a BPS crust \cite{bps} to the EoS. The MBF model used here contains $\omega\rho$ contributions (coupling strength of $40$), following the same motivation as for the CMF model. The hybrid stars are constructed in this case via a Maxwell construction leading to a quark phase in which the higher-order vector self interaction $\omega^4$ term is also included with strength $23$ (for the first time in this formalism), while the Bag constant is B=77 MeV fm$^{-3}$ and the quark vector coupling is $a_0 = 3$ fm$^2$ \cite{Gomes:2018bpw,Gomes:2018eiv}. 

The respective particle population is shown in the top panel of Fig.~5. Due to the early deconfinement to quark matter, only the Lambda hyperons appear in the hadronic phase. In the quark phase, the amount of strange quarks increases with density. The stellar masses and radii generated by this hybrid EoS are shown in Fig.~6 and Tab.~III, where we also show the pure MBF hadronic version for comparison.

In a completely different approach, we describe pure quark stars using the vector interaction enhanced bag model (vBag) \cite{Klahn:2015mfa}, which is parametrized to reproduce absolutely stable strange matter \cite{Salinas:2019fmu} with a bag constant B=59 MeV fm$^{-3}$ and quark vector coupling K$_v= 70$ fm$^2$. It is further worth mentioning that the main difference between vBag and standard Nambu–Jona-Lasinio (NJL) models in the chirally restored phase is the value of the effective vacuum bag constant, which due to confinement is expected to be smaller in bag type models and, hence, more likely in support of the strange matter hypothesis \cite{Klahn:2015mfa}. As can be seen in Fig.~6 and Tab.~III, vector repulsion is essential and sufficient to model high-mass pure quark stars. The respective particle population is shown in the bottom panel of Fig.~5. At low densities, matter is more isospin asymmetric and there are no strange quarks. At large densities, there is about the same number of each quark.

\begin{table}[t!]
\begin{tabular}{c|ccccc}
~Interactions~    &~$M$($\Msun$)~&~$n_c(n_0)$~&~$v_s^2$~&~$R_{1.4}$~(km)~&~$\Lambda_{1.4}$~\\ 
\hline
  MBF         &  2.14  &  5.8  &  0.51   &  13.6   &  823  \\
  MBF + vBag  &  2.14  &  5.8  &  0.40   &  13.6   &  823  \\
  vBag        &  2.26  &  5.8  &  0.47   &  11.1   &  144  \\
\end{tabular}
\begin{center}
\caption{Same as Table I but for other equation of state models.}  
\end{center}
\label{tab2}
\end{table}

\begin{figure}[t!]
\centering
\includegraphics[{trim=0.5cm 0 0 2.4cm},clip,width=9.4 cm]{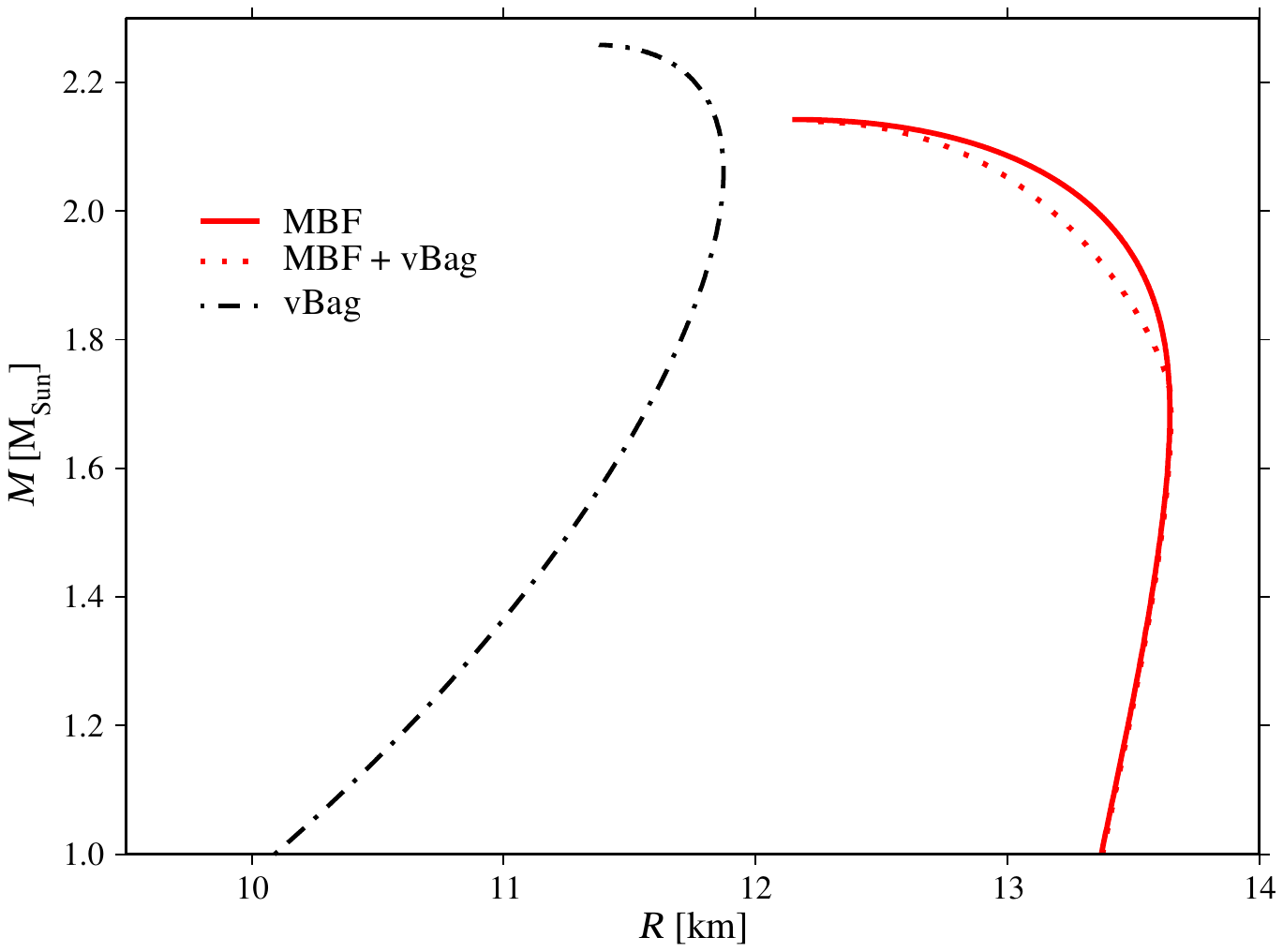}
\caption {Static mass-radius diagram for different equation of state models (shown until the maximum mass only).}  
\label{fig3}
\end{figure}

Figure 7 shows the impact of rotation effects on the two additional models discussed above. The largest possible gravitational mass found for stars increases again with increasing rotational frequency, easily achieving and going above $2.5\,\Msun$. Once more, for the combined model, an increase in rotation at fixed (or small increase in) baryon number decreases the stellar central density, suppressing quark but not hyperonic degrees of freedom. This is not the case again for stars that rotate fast but possess a large amount of baryons (than what is allowed in the static case). Once more, the upper-right corner of the top panel of Fig.~7 shows that stars with about $0.4\,\Msun$ extra baryon mass (than the static maximum mass) can have masses $\sim 2.6\,\Msun$ and still contain exotic degrees of freedom, namely hyperons and quarks.

In this case, the stability criterion does not exclude any kind of exotic degrees of freedom due to the very early setup of deconfinement, although the total amount of strangeness is reduced by allowing for a lower amount of s-quarks. The characteristics of the most massive maximally spinning configuration are shown for this combination in the third line of Tab.~II.

\begin{figure}[t!]
\includegraphics[width=0.48\textwidth]{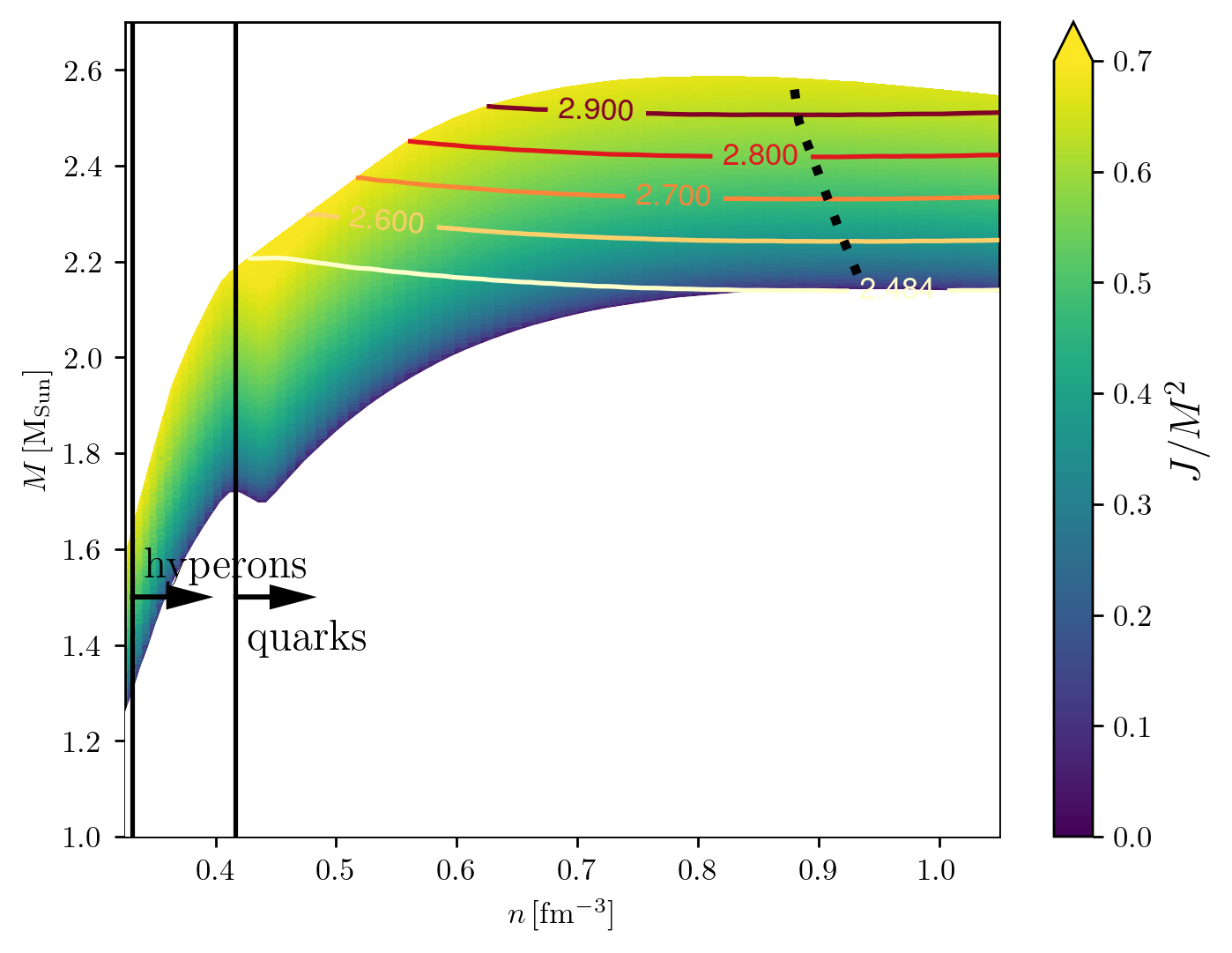}
\includegraphics[width=0.48\textwidth]{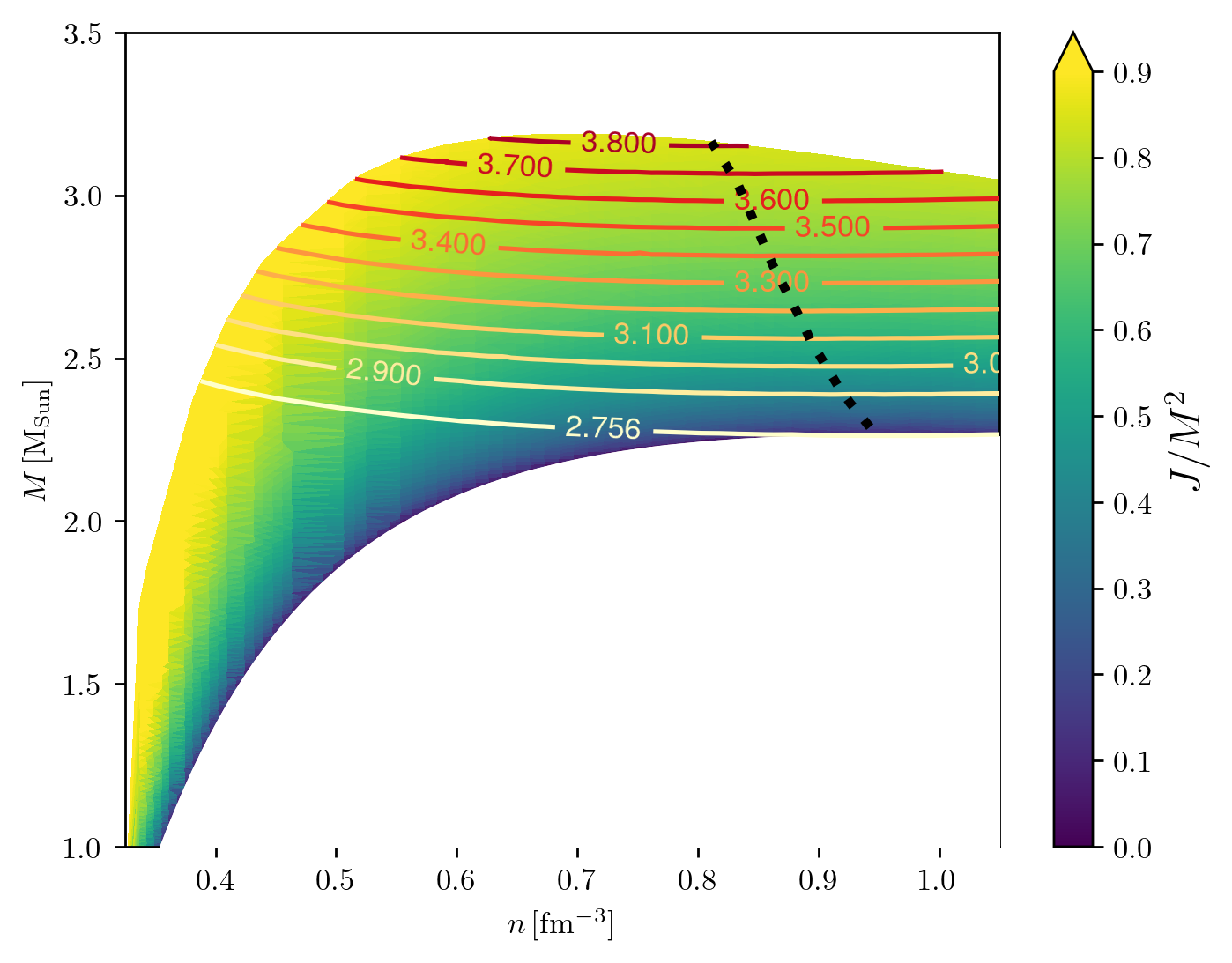}
\caption {Mass-central density diagram for rotating stars reproduced by different equation of state models: the MBF+vBag combination in the top panel and the vBag in the bottom one. Labels are the same as Fig.~4.}
\label{fig4}
\end{figure}

Only in the pure quark model, are the quarks never suppressed when fast rotation is included, irrespective of the increase or lack of increase in baryon number when compared to the static solution. But, in particular, an increase in mass at fixed baryonic number (the maximum static one), which decreases the stellar central density from $n_B=0.9\to0.4$ fm$^{-3}$, modifies the particle composition by replacing some of the strange quarks by down quarks, decreasing the total stellar strangeness content. But this model also predicts much more massive fast rotating stars that possess $\sim1\,\Msun$ of extra baryon mass (than the static maximum mass). The characteristics of this most massive maximally spinning configuration are shown in the fourth line of Tab.~II. Note that all frequency limits in this table are within the range predicted by Ref. [10], which used universal relations to generate rapidly-rotating massive stars.

\section{Conclusions and Discussions}

We have presented in this work an extension of the Chiral Mean Field (CMF) hadronic and quark relativistic mean-field model that contains new vector interactions. This model stands out, as it was formulated to be applicable in different regimes, including hot temperatures produced in relativistic heavy-ion collisions and intermediate temperatures produced in neutron-star mergers and supernova explosions. The novelty lies in the vector interactions included showing for the first time the particle population changes caused by the $\omega\rho$ self-vector interaction, together with new terms for the quark-meson interaction, and $\omega^4$ and $\omega^6$ self-vector interactions. We have also modified the deconfinement order parameter potential $U$ in order to generate hybrid stars with a stable pure quark core, without the need of generating mixtures of phases (as the standard CMF parametrization). Although it has been shown that strong vector interactions are not favored by lattice QCD \cite{Ferroni:2010xf,Steinheimer:2010sp}, those calculations are performed in the large temperature, low density limit and do not necessarily extrapolate to our low temperature, high density regime. 

We have also made comparisons with other two relativistic descriptions, the Many-body Forces (MBF) model \cite{Gomes:2014aka}, which describes the interaction of baryons considering higher-order scalar self-couplings interpreted as many-body forces, combined with a vector-enhanced Bag (vBag) model for the quarks and the vector interaction enhanced bag (vBag) model alone. We demonstrated that in all cases vector interactions can increase the maximum mass of stable non-rotating hybrid stars to $\simeq 2.1\,\Msun$, but at the same time not creating conflict with other observations of for example neutron star radii and tidal deformabilities. This is because the vector interactions stiffen the equation of state of nuclear matter mainly at intermediate and large densities (or chemical potentials) and we combined it with a $\omega\rho$ interaction that tends to soften the EoS at low and intermediate densities.

For each equation of state, we have calculated using a full general-relativity numerical code the parameter space of rapidly rotating neutron stars up to break-up frequencies of $\simeq 1.5\, \rm{kHz}$. We found that even when rotating this fast, both hybrid EoSs still can reproduce stars that contain quark degrees of freedom with masses $\simeq 2.5\,\Msun$ (for some parametrizations), but only if they possess considerably more baryons than what is allowed for static stars. In this case, they cannot survive as neutron stars as they spin down and must, eventually, collapse to black holes. Such rapidly rotating short lived neutron stars could be formed, for example, through low-mass binary neutron star mergers \cite{Giacomazzo2013}.

But, no matter the initial stellar mass and baryon number, as they spin down at fixed baryon number, their central densities increase and (unless their masses are too small) surpass the threshold for deconfinement. This kind of event might have interesting observables related to, for example, oscillations \cite{Abdikamalov:2008df,Staff:2011zn} and ejecta \cite{Ouyed_2002}. Note that several works in the literature have already investigated how the appearance of quarks in supernova events can trigger the explosion of massive stars \cite{Sagert:2008ka,Fischer:2017lag,Zha:2020gjw}, but much more work is needed to fully understand this scenario for different gravitational and also different baryon masses. Work on this topic using the CMF model is in progress.

The comparison with the pure quark model is a little bit more interesting, as in this case quarks are obviously  present in all possible rotating configurations. Nevertheless, it is interesting that this EoS also allows $\sim 2.5\,\Msun$ stars to have approximately the same amount of baryons as non-spinning solutions. This means that such stars could spin down without collapsing to a black hole. Of course, this feature is related to the larger freedom in modelling quark models, which do not have to reproduce nuclear saturation properties.

We highlight the importance of studying the effects of rotation on stars that contain exotic matter as a way to learn about dense matter and, more specific, nuclear interactions. For example, different self-coupling vector interactions that reproduce about the same static maximum mass stars behave differently under fast rotation, only in one of the cases (that generates stiffer equations of state at all densities) it allows for sizable stable hybrid stars with pure quark cores rotating at the limiting frequencies. In addition, the pure quark model that generates about the about the same static maximum stellar mass (as the hybrid models) generates much more massive stars with a much larger baryon number and much larger angular momenta when rotating at limiting frequencies. Nevertheless, all models presented a decrease in strangeness for stars that rotate faster.

We finally comment on the implications of our work for a hypothetical neutron star origin of the secondary object in GW190814. We have shown that rapid rotation can increase the maximally allowed mass of rotating hybrid stars to $\simeq 2.5\,\Msun$. This indicates that, with further refinement of the models, having hyperons and deconfined quarks is not at odds with supporting a massive neutron star, while also not being in conflict with other recent astrophysical observations of stellar radius and tidal deformability. This type of analysis allows us to learn more about the attractive and repulsive components of strongly interacting matter in a regime otherwise not accessible. Furthermore, we find that even a slowly rotating pure quark star could explain a non-black hole massive secondary object. This is a feature that is beyond the mass range comfortably attainable with realistic hybrid-star models.

Note that around and after our work appeared online, other works showed that hybrid stars could explain the secondary object in  GW190814 using alternatively a holographic description for quark matter \cite{Demircik:2020jkc}  or large  constant speed of sound description with $v_S=c$ \cite{Christian:2020xwz}, that pure quark stars could explain it through the two-flavor hypothesis \cite{Cao:2020zxi}, using large CFL gaps $\Delta>200$ MeV \cite{Roupas:2020nua}, or invoking electric charge separation \cite{Goncalves:2020joq}, or finally that there is another 2-family approach to the problem \cite{Bombaci:2020vgw}.

\section*{Acknowledgements}
We thank Agnieszka Sorensen for useful discussions.
Support for this research comes from the National Science Foundation under grant PHY-1748621 and PHAROS (COST Action CA16214). S.H. acknowledges support from the National Science Foundation, Grant PHY-1630782, and the Heising-Simons Foundation, Grant 2017-228.

\bibliographystyle{apsrev4-1}
\bibliography{paper}
\end{document}